# Integrating Graphene into Semiconductor Fabrication Lines

Daniel Neumaier[1], Stephan Pindl[2], Max C. Lemme[1,3]

[1]AMO GmbH, Otto-Blumenthal-Straße 25, 52074 Aachen, Germany
[2]Infineon Technologies AG, Wernerwerkstraße 2, 93049 Regensburg, Germany
[3]Chair for Electronic Devices, RWTH Aachen University, 52062 Aachen, Germany

*Electronic and photonic devices based on the two-dimensional material graphene have unique properties, leading to outstanding performance figures-of-merit. Mastering the integration of this new and unconventional material into an established semiconductor fabrication line represents a critical step for pushing it forward towards commercialization.*

Silicon has remained the dominating material in microelectronics for more than five decades. Although many semiconductors like Ge, GaAs or InP possess higher charge carrier mobilities, favourable optical properties or other advantages compared to silicon, the simpler production and processing of the latter make it by far the most convenient and cost effective material for large markets.

This lesson applies equally well to graphene, the two-dimensional allotrope of carbon that moved into the focus of research [1] because of several outstanding intrinsic properties, such as a very high carrier mobility [2], broadband optical absorption covering the far-infrared (IR) to ultra-violet (UV) range [3] or extremely high surface to volume ratio. On a single device level these intrinsic properties have already been utilized for realizing high performing devices, including infrared photodetectors [4,5], Hall-effect magnetic field sensors [6], pressure sensors [7] and gas sensors [8], which all outperform their counterparts based on established semiconductors. Sensing may therefore be a promising application space; yet entry to this market — currently dominated by silicon and silicon MEMS (**m**icro**e**lectro**m**echanical **s**ystems) technology for magnetic field, pressure or gas sensors in large volume automotive and consumer electronics, and by other more expensive semiconductors like InGaAs for infrared detectors and imaging systems in smaller markets such as scientific instrumentation, security or defence — is still hampered by the lack of a scalable device manufacturing process. The development of a reliable large-scale production process may not only unleash the potential of graphene in the sensors competition, but may also help triggering other key applications, where the unique properties of this material make a difference.

In this respect, graphene integration into conventional silicon-based fabrication lines can be a promising direction, as it allows banking on widespread and well-established processing steps with relatively low engineering effort compared to the development of entirely new production lines from scratch. In particular, 3D integration on top of the silicon CMOS (Complementary Metal-Oxide-Semiconductor) platform may enable the combination of high performance graphene devices with established CMOS readout circuitry at production costs similarly low as for conventional silicon technology. The first proof of concept demonstrations of such 3D integrated graphene sensors systems were magnetic field sensors [9] and infrared image sensor arrays [10].

Beyond graphene, other layered two-dimensional materials like transition metal dichalcogenides (TMDCs), black phosphorus and hexagonal boron nitride (hBN) as well as their heterostructures have also raised great interest in basic and applied research due to their intriguing properties [11]. These 2D materials have different application scenarios than graphene, as shown in Fig.1, but face similar challenges related to growth, processing and integration into semiconductor fabrication lines. Thus graphene can be considered as a model system for 2D materials when it comes to semiconductor fab integration, and the lessons learned from graphene may directly be applied to other 2D materials.

Here we provide an outline on how the integration of graphene can be managed, using a silicon CMOS platform as an exemplary case study. We notice that the specific process steps and the integration scheme will differ depending on the specific application technology developed — for instance, membrane based pressure sensors or microphones are expected to require different fabrication processes than infrared photodetectors. Such a diversity in the manufacturing process is also known as the "MEMS law" for the case of silicon MEMS technology, where nearly every product requires a specific and unique fabrication technology. For the sake of generality, we will focus our discussion on the major optimization challenges occurring in the most common microelectronics fabrication steps and their possible solutions.

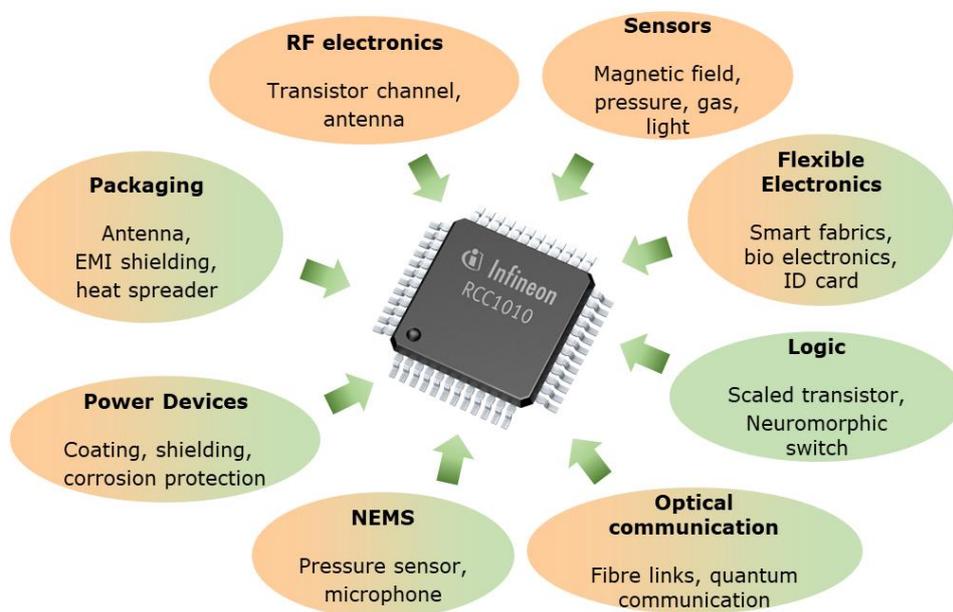

**Fig. 1 Applications for two-dimensional materials.** Potential application fields for graphene and related two-dimensional materials where wafer-scale integration is required. Applications addressed mainly by graphene are coloured light red, and applications addressed by TMDCs are coloured green.

**Front end of line vs. back end of line integration**

Semiconductor manufacturing [12] is typically separated into the front end of line (FEOL) and the back end of line (BEOL), which not only defines the status in the production line, but also sets the boundary conditions for the process steps involved and thus affects how new materials can be integrated into the entire process. In general, FEOL includes the first steps in integrated circuit fabrication, which are mainly related to transistor / device fabrication. BEOL processing essentially involves the fabrication of the metal interconnects, corresponding dielectric layers and diffusion barriers. A schematic cross-sectional view of a typical silicon CMOS structure indicating the FEOL and BEOL parts is shown in figure 2a. This distinction and the choice between FEOL and BEOL is very important, as it defines the process parameter space for the integration.

If graphene is integrated during the BEOL steps, it will be relatively far away from the active silicon devices. At that level, there are significantly fewer restrictions regarding metal contamination compared to the FEOL, as integrated diffusion barriers prevent damage to the silicon devices. Furthermore, the typical processing temperatures reached during the BEOL stages are relatively modest, below ~450 °C or even below ~150 °C, meaning that thermal stress and corresponding degradation is expected to be minimal. However, this implies that the integration of graphene may not involve any high temperature process steps, which damages the underlying layers. One should also keep in mind that the materials in the BEOL are not crystalline, but with the surfaces being of amorphous nature or consisting of polymers. The absence of crystalline surfaces and the strict temperature limit are the main reasons why BEOL-compatible growth of semiconductors like Ge, GaAs or InP is not available. This is instead a crucial advantage and opportunity for graphene, as it can be grown on amorphous surfaces or transferred on them after being grown on a separate substrate.

In contrast, at the FEOL, when the transistors are built, all integrated materials need to sustain high temperatures, as dopant activation requires heating the wafers up to ~1000 °C. In addition, the integration of new materials must not introduce contamination especially from metals like copper, gold or silver, which are highly mobile atoms in silicon and cause deep trap states that affect the performance of transistors [12]. This is an additional challenge for graphene transferred from metallic growth substrates, as complete absence of any potential contaminant from its surface must be ensured.

Integration of graphene devices with silicon CMOS logic circuits that can be used for controlling, data readout and data processing, require an all-BEOL integration scheme (Figs.2b-f) to avoid affecting the silicon transistors fabrication and performance thereof. Hence in the following sections we will focus the discussion on BEOL integration as one possible examples. We expect that different applications will require different solutions for the involved process steps, but the basic problems and challenges will be very similar and thus the discussion can be applied in a broader context.

**Current status and challenges**

The first step for graphene integration is its growth which, as mentioned above, can directly occur on the target surface, or performed on a separate substrate with subsequent transfer to the target surface.

Direct growth of graphene on an amorphous surface at BEOL-compatible temperatures has been demonstrated, for instance by plasma-enhanced chemical vapour deposition (PECVD) [13], but the resulting layers are quite defective. As such, devices using these materials will hardly provide outstanding performance. Another option is the deposition of a metal catalyst layer on top of the substrate, followed by graphene growth and subsequent metal removal [14]. However, this method requires high temperatures, that are not compatible with BEOL.

Graphene grown on a separate template and then transferred to the target surface is a very elegant solution, as it decouples the growth process from the final substrate. Consequently, high temperatures may be used for growth, and the underlying substrate may be optimized with respect to catalytic activity and crystalline orientation, which are the main parameters defining the graphene quality. The CVD growth of graphene on copper and platinum [15] is well developed and scalable to a size compatible with state-of-the-art Si technology (300 mm wafers), or even larger. Carrier mobilities exceeding 100.000 cm²/Vs have been reported at room temperature for CVD grown single crystalline graphene islands [16], while polycrystalline layers reach lower mobility values on the order of ~10.000 cm²/Vs [17]. A promising combination of both methods is the growth of merged single crystalline graphene islands having the same orientation, leading to quasi single crystalline and continuous films with carrier mobility up to ~15.000 cm²/Vs [18,19].

While the CVD growth of graphene on metal surfaces is well developed and enables carrier mobilities close to the theoretical phonon limit, the transfer process is presently the limiting step for the device performance. Transfer includes delaminating graphene from the metal, handling and depositing it on a silicon wafer. The available wafer scale transfer processes [20] use either polymers to handle the graphene and/or wet chemistry for releasing the graphene from the metal surface. This typically leads to a contamination from polymer residues or other involved chemicals, degrading device performance. Even though substantial efforts have been invested to improve post-process cleaning, it is challenging to completely restore graphene performance to intrinsic values. Conventional cleaning methods such as oxygen plasma exposure cannot be applied, as this would etch graphene entirely. In addition, the handling of the graphene layer during transfer can introduce mechanical damage or excessive stress, which also reduce device performance or even cause complete device failure.

The highest mobility values for CVD grown graphene have been obtained by mechanically delaminating it from an oxidized copper surface via van- der- Waals forces with a hBN flake and placing it onto another hBN flake [16], taking advantage of their atomically flat and clean surfaces. With this van- der- Waals bonding transfer several degradation mechanisms are avoided: graphene is not exposed to any chemicals like acids, $H_2O$ or polymers, so that its surface is kept clean. In addition, mechanical damage or stress during transfer is suppressed. The main limitation of the van- der- Waals bonding process is the small size of the transferred graphene, typically on the order of 100 µm. This limit is set by the size of the single graphene crystals grown on copper surfaces and, even more so, by the size of the hBN flakes. While templated growth of single crystals at predefined locations or quasi-single crystalline growth offers a workaround to the size limitation for graphene, it is desirable yet challenging to replace micromechanically exfoliated micron-scale hBN by large area counterparts. In addition, it is still unclear how this mechanical delamination of graphene from the metal surface would work on wafer scale.

Dielectric interfaces and encapsulation are further integration aspects that have not been solved on wafer scale yet. The above-described strategy leading to a graphene layer embedded between two multilayer hBN dielectrics layers at present results in devices with the highest performance in terms of mobility and residual charge carrier concentration, but wafer-scale van-der-Waals substrates or thin films with similar quality are not available yet. Alternatively, conventional oxides or nitrides, which can be grown on wafer scale, are also widely studied as dielectrics. These systems can be used in pristine or functionalised form, with the optimal amount of covalent bonds at the interface to graphene resulting from a trade-off between the electronic performance of graphene and its adhesion to the substrate, which is key to prohibit delamination during subsequent processing. Carrier mobilities in graphene around ~10.000 cm²/Vs have been demonstrated using large-scale transfer on conventional surfaces like $SiO_2$ [19]. Even though this is still far below the theoretical and experimental record values for CVD graphene, it is sufficient for several promising device applications, giving outstanding performance in infrared photodetectors, pressure sensors, gas sensors or electro optical modulators.

Electrical contacts to graphene are essential for any electronic, photonic and sensor device [21]. As a general rule, the contact resistance is considered acceptable if it does not significantly contribute to the total device resistance. For example, a typical Hall sensor has an intrinsic resistance of 2 kΩ and the contact leads are 20-100 µm wide. This means that a width-specific contact resistance below 1 kΩµm would be sufficient as it would contribute less than 5% to the total device resistance. For sub-micrometre scale devices like transistors or electro-optical modulators, the specific contact resistivity should be 100 Ωµm or less in order to fully exploit their performance potential. In the past years significant efforts have been devoted to develop low-ohmic contacts to graphene, and there are several options available that provide sufficiently low contact resistances for most applications. The most straight-forward option is depositing a metal on top of the graphene (top-contact configuration), which has led to with specific contact resistances down to 50-100 Ωµm [21]. However, if the graphene is encapsulated before the contact fabrication, contact holes are required. This is challenging, because stopping the etching process on top of monoatomic graphene layers is nearly impossible. Thus one-dimensional edge contacts have been developed for encapsulated devices, which provide specific contact resistances on the order of 200-300 Ωμm [2]. The lowest contact resistance so far has been achieved with a combination of edge and top contacts, that is top contacts on perforated graphene, with values down to 23 Ωμm [22]. The choice of contact metal is also crucial for the contact resistance, and there are only certain metals available that are compatible with silicon technology including Al, W, Cu, Ni, Ti and Ta. Good electrical contacts for graphene have been obtained with Ni or Ti [21], two metals that are readily available in a conventional fab. Other metals like Au or Pt could generally be introduced at the BEOL, but this would require changes in the standard line. In contrast to graphene, ohmic contacts to semiconducting TMDCs are still an open issue. In fact, most metals form non-ohmic Schottky junctions to the TMDC layer, resulting in relatively high and bias-dependent contact resistances [23]. This problem is well known in silicon technology and the solution is heavy doping of the silicon in the contact area. However, such doping techniques cannot be directly applied to TMDCs and thus further research will be needed to solve the contact problem in these materials. An alternative route for the formation of ohmic contacts to TMDCs may be offered by controlled phase transformation from the semiconducting to the metallic phase [24].

| **Step** | **Critical Parameter** | **Possible solutions / approaches** |
|---|---|---|
| **Growth** | Nucleation density | Seeded growth using predefined nucleation points; reduction of process gas pressure and/or concentration. |
| | Defect Density | Reduction of process gas flow; optimal growth temperature; specific $O_2$ gas flow. |
| **Transfer** | Delamination from growth substrate | Water or ion intercalation and mechanical peeling off; removal of growth substrate by wet-chemical etching. |
| | Handling during transfer | Coating with handling polymer; lamination with handling foil (e.g. thermal release tape); deposition of (van-der-Waals) dielectric on top; using sacrificial layer in-between polymer (or foil) and graphene. |
| | Removal of handling material | Using wet-chemical solvents (in case of polymers) plus removal of any sacrificial layer wet-chemically; using the top (van-der-Waals) dielectric as functional layer in the device; releasing the foil by light, temperature or other method. |
| **Dielectric Environment** | Substrate surface | Polishing; defined surface termination using functionalization (e.g. oxygen plasma, deposition of self assembled monolayers); pure van-der-Waals substrate surface (e.g. hBN). |
| | Adhesion on Substrate | Pure van-der-Waals substrate surface (e.g. hBN); defects in the graphene (at predefined locations) for $sp^3$ bonds; partial graphene coverage and clamping by contacts or encapsulation. |
| | Interfacial control | Graphene lamination under controlled environment (e.g. vacuum); in-situ substrate functionalization. |

| | Deposition of dielectric on top | Graphene surface functionalization followed by ALD; deposition of seed layer followed by ALD; lamination of van-der-Waals dielectric (e.g. hBN); direct deposition of polymeric dielectric. |
|---|---|---|
| **Electric Contacts** | Metal deposition | Deposition on-top of graphene avoiding interfacial contamination; one-dimensional edge contacts; sandwich contacts avoiding interfacial contamination; combination of these contacts schemes. |
| | Work function control | Proper selection of the metal (e.g. Ni, Au, Pd, etc.). |

Table 1: Critical process steps, parameters and possible solutions for the integration of graphene into a semiconductor manufacturing line.

In general, solutions exist for all major process steps required for the integration of graphene with a BEOL compatible process flow, although none are ready for production today. Table 1 summarizes these main steps, their critical parameters and potential technological solutions. As previously discussed the most critical ones for the final device performance are related to the graphene growth, the transfer process and the dielectric interface of the graphene. While for the graphene growth there are already very promising solutions available, which enable mobilities close to the intrinsic limits, the transfer process and the dielectric interfaces are the major challenges to be solve in order to unlock the full performance of graphene based devices. Ideally the transfer process should completely avoid wet chemistry and polymers, while the dielectric interfaces to the graphene layer need to be as inert and as smooth as possible. In contrast, electrical contacts are already rather well developed and sufficient solutions are available at least for micron scale devices.

An example of a feasible process flow for the BEOL integration on a typical silicon CMOS structure is illustrated in figure 2. Such a process could be used e.g. for the fabrication of graphene-based Hall-Sensors or similar devices. The main challenges ahead towards applications are device reproducibility, fabrication yield and reliability. While these aspects are not in the focus at an early research stage where hero devices drive progress, large-scale production absolutely depends on statistical data on the above. The low reproducibility and rather low yield of graphene devices originate mostly from manual or semi-automated handling, which is significantly less reproducible compared to fully automated manufacturing. Thus moving on to a fully automated process line will be a big step ahead in this respect. The challenge here is the transfer of the graphene to the target substrate, for which new tools need to be developed. In parallel, device stability and reliability need to be studied in-depth, including wafer scale device measurements and data analysis. Investigations of the relevant failure mechanisms are required to understand how to provide stable operation during the typical life-time of a product, which may be several years for consumer electronics or more than 20 years for automotive applications.

System design with graphene is currently a chicken-egg problem. There are plenty of device models that allow in principle designing circuits and systems [25], but, as discussed, devices lack the reproducibility and stability. Thus, early design concepts cannot be generalized and standardized. We propose to employ a material-device-circuit co-design approach [26] in which experts in materials synthesis and device fabrication engage with the electronic circuits and systems communities to take into account circuit and system level requirements since the earliest stage of technology development.

Feedback loops need to be established that enable material and design changes based on circuit-level figures of merit.

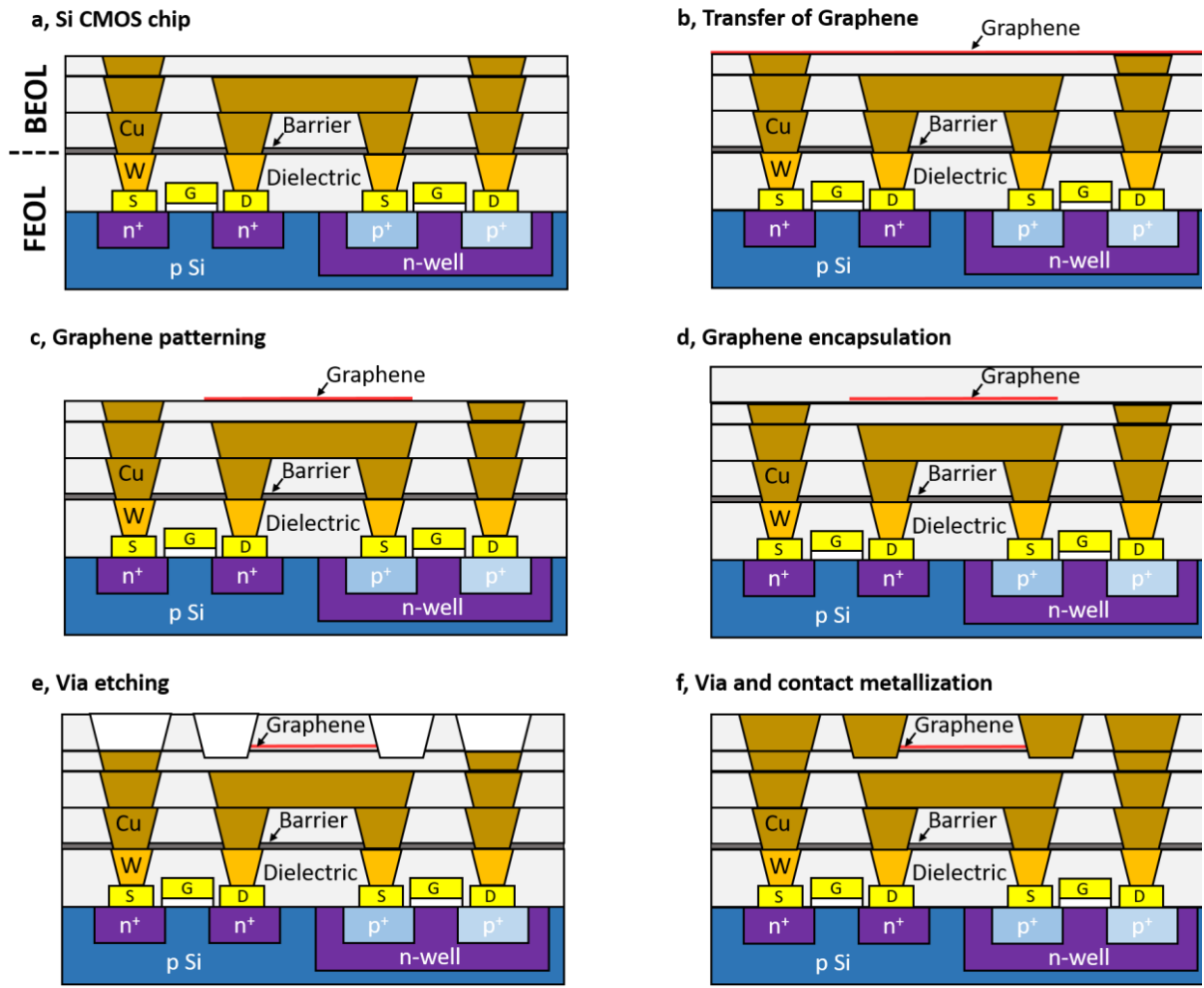

**Fig. 2 Flow chart for BEOL integration. a**, Cross sectional schematic of the CMOS chip before graphene integration, indicating the FEOL and BEOL parts. **b**, Transfer of the graphene onto the entire wafer. **c**, Patterning of the graphene layer using oxygen plasma. **d**, Encapsulation of the graphene using a dielectric layer. **e**, Via etching through the top dielectric layer by means of dry etching techniques. **f**, Filling of the vias by metal providing edge contacts to the graphene. After step (**f**) further interconnect layers can be fabricated in order to connect the graphene layer to the Si CMOS devices. This integration schematic can be used e.g. for integrating graphene Hall-Sensors on wafers containing Si CMOS logic.

**Outlook**

Compared to other semiconductors like Ge, GaAs or InP, graphene offers the key advantage of being compatible with BEOL processing and integration. This provides a unique opportunity to extend the functionality of silicon CMOS circuits with the integration of different electronic, photonic or sensor devices based on graphene, without requiring a compromise or significant changes at the CMOS level. In addition, the basic process steps can be adapted for the specific needs of other application technologies. Even though challenges remain, there is no fundamental road-block towards the waver-scale processing of graphene devices. However, engineering problems such as reproducibility, variability, fabrication yield and durability of the devices must be addressed. A major step towards statistically relevant data sets is expected to happen, once processing is performed on wafer scale and fully automated. Afterwards, improvements will result from continuous learning processes, similar to what happened in the last five decades in silicon technology. It is further expected that graphene devices will not suddenly pop up as performance boosters in CMOS systems, but that the market penetration will start from niche applications like the quantum Hall resistance standard [27], where standalone graphene devices offer a unique and significant advantage. Such niche applications can already be addressed with the current state of production and will allow an organic growth of the whole ecosystem. Next, medium sized markets are expected to emerge, where devices are no longer manually manufactured and hand-selected, yet production costs will still be higher than stand-alone silicon-based systems; these will however be justified by boosts in functionality obtained thanks to the use of graphene. Examples of these developments could be IR imaging systems or ultra-high speed optical communication links. We expect this second market penetration to happen in the upcoming two to eight years. Finally, once a basic ecosystem and supply chain are established, large volume production can be expected. Nevertheless, it is still rather early to predict which graphene-based device will make it there first, or if other 2D materials will surpass graphene in that respect.

**Acknowledgements**

All authors acknowledge funding from the European Union H2020 Graphene Flagship Project (grant agreement No. 785219).

**Authors contributions**

All authors conceived this work and collaborated equally in the writing of the text.

**Competing interests**

The authors declare no competing interests.